\documentclass{jpsj3}

\usepackage{color}

\def\gsim{\mathop {\vtop {\ialign {##\crcr 
$\hfil \displaystyle {>}\hfil $\crcr \noalign {\kern1pt \nointerlineskip } 
$\,\sim$ \crcr \noalign {\kern1pt}}}}\limits}
\def\lsim{\mathop {\vtop {\ialign {##\crcr 
$\hfil \displaystyle {<}\hfil $\crcr \noalign {\kern1pt \nointerlineskip } 
$\,\,\sim$ \crcr \noalign {\kern1pt}}}}\limits}

\title{
On Sharp Enhancement of Effective Mass of Quasiparticles and Coefficient of $T^{2}$ Term of Resistivity 
around First-Order Metamagnetic Transition
Observed in UTe$_2$
}

\author{
Kazumasa Miyake 
}
\inst{
Center for Advanced High Magnetic Field Science, Osaka University, Toyonaka, 
Osaka 560-0043, Japan 
}

\recdate
{
July 7, 2020
}

\abst
{
The mechanism underlying the enhancement of the Sommerfeld coefficient  of quasiparticles 
at the {\it first-order} metamagnetic transition in UTe$_2$, reported by  
Miyake {\it et al.} in J. Phys. Soc. Jpn. ${\boldsymbol  88}$, 063706 (2019), is discussed theoretically by 
taking into account the ferromagnetic order-parameter fluctuations on the basis of the Landau theory of 
phase transition.  We find that the enhanced ferromagnetic spin fluctuation gives rise to 
the enhancement of the effective mass of the quasiparticles or the Sommerfeld coefficient  $\gamma$, 
which is  consistent with the experimental observations. At the same time, 
the Kadowaki-Woods type scaling around the metamagnetic transition, reported by 
Imajo {\it et al.} in J. Phys. Soc. Jpn. ${\boldsymbol  88}$, 083705 (2019) and 
Knafo {\it et al.} in J. Phys. Soc. Jpn. ${\boldsymbol  88}$, 063705 (2019), is also understood 
semiquantitatively by assuming reasonable values of parameters of Landau-type free energy reproducing key 
quantities characterizing the metamagnetic transition. 
}

\begin{document}
\sloppy
\maketitle

%%%%%%%%%%%%%%%%%%%%%%%%%%%%%%%%%%%%%%%%%%%%%%%%%%%%%%%%%%%%%%%%%%%%%%%%%%%%%%%%%%
\section{Introduction}
The discovery of superconductivity in UTe$_2$ reported in December 2018~\cite{Ran} gave a 
strong impact in the heavy fermion community, and it was confirmed soon after by a subsequent 
experiment,~\cite{Aoki1} in which a number of intriguing physical properties other than the superconductivity 
were reported.  In particular, unusual metamagnetic behaviors have attracted considerable 
attention.~\cite{AtsushiMiyake, Knafo, Ran2 }  
Among them, it is a non-trivial phenomenon that the Sommerfeld coefficient  and the $A_{\rho}$ coefficient of 
the $T^{2}$ term in the resistivity exhibit sharp enhancements around the {\it first-order} 
metamagnetic transition.~\cite{AtsushiMiyake,Imajo,Knafo} 
UTe$_2$ is considered to be located in the normal phase near the ferromagnetic quantum critical point 
under the ambient conditions.~\cite{Ran,Aoki1,Sundar}  
Therefore, it is natural to expect that such anomalous properties arise through the effect of ferromagnetic 
spin fluctuations of one kind or another, which is enhanced by the {\it first-order} metamagnetic transition.  
It is remarked that this first-order metamagnetic transition should be conceptually different from the first-order 
ferromagnetic transition with the tricritical wings in the $T-P-H$ phase diagram, which was discussed 
in Ref.\ \citen{Kirkpatrick}.    

In this paper, the origin of such a non-trivial behavior is clarified on the basis of 
an extended phenomenological theory 
of the Landau type and the conventional theory of ferromagnetic spin fluctuations.  We are not so ambitious to explain 
the magnetic field dependence at an arbitrary magnetic field strength and its anisotropic behaviors with respect to 
the direction of the magnetic field, but to focus on the physical properties only at exactly the metamagnetic 
field and ambient pressure in a region with sufficiently low temperatures on the basis of a phenomenological 
uniaxial model for the magnetization, i.e., in the $b$-direction, as observed in UTe$_2$. 
The use of this phenomenological model may be justified by the fact that the direction of the magnetic field 
($b$-direction in UTe$_2$), in which metamagnetic transition occurs, and that corresponding to the maximum of 
the magnetic susceptibility  ($a$-direction in UTe$_2$) are different in general as discussed in Sect. 2.

This paper is organized as follows. In Sect. 2, a theory for the {\it first-order} metamagnetic transition 
is formulated on the basis of an extended Landau theory of phase transition.  
On this basis, in Subsec. 3.1, the structure of ferromagnetic spin fluctuations around the {\it first-order} 
metamagnetic transition is presented, which shows that such enhancements in the effective mass and 
the $A_{\rho}$ coefficient are caused by effect of the flattening of the curvature of the free energy $F(M)$ around the local minimum corresponding to the lower magnetization 
at the  {\it first-order} transition. It is shown that such an effect is manifested also in the $A_{\rho}$ coefficient. 
On this consideration, in Subsect. 3.2, the observed manner of enhancements of the Sommerfeld coefficient  
and the $A_{\rho}$ coefficient is evaluated semiquantitatively. 
Throughout this paper, we use units of energy, such that $\hbar=1$, $k_{\rm B}=1$, and $\mu_{\rm B}=1$, 
unless explicitly stated. 
In Sect. 4, the results of this study are summarized, and their relevance to the re-entrant appearance of 
the superconductivity at $B\gsim B_{\rm m}$ is briefly mentioned. 
%In particular, the apparent breaking of the Kadowaki-Woods ratio is understood naturally.  

%formulas for the enhancement of the effective mass and the $A_{\rho}$ coefficient are derived  

%A less ambitious but fundamental problem concerning the interesting phenomena associated 
%with the metamagnetic transition observed in UTe$_2$, which is a newly discovered U-based 
%superconductor, is discussed theoretically.  

%%%%%%%%%%%%%%%%%%%%%%%%%%%%%%%%%%%%%%%%%%%%%%%%%%%%%%%%%%%%%%%%%%%%%%%%%%%%%%%%%%
\section{Extended Landau Theory for Metamagnetic Transition}
\begin{figure}[h]
\begin{center}
\rotatebox{0}{\includegraphics[width=0.7\linewidth]{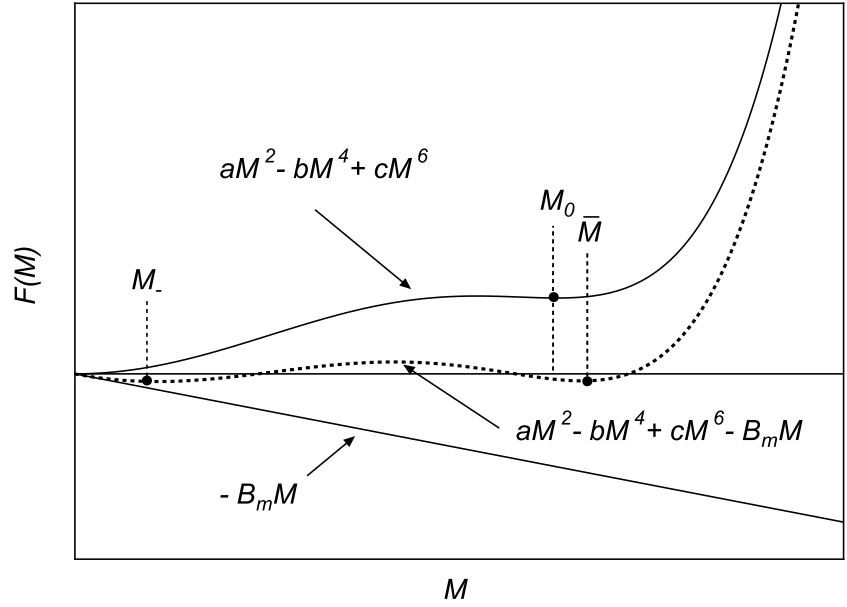}}
\caption
{Schematic behavior of free energies as a function of the uniform magnetization. The solid curve and solid line 
represent the free energy $F_{0}(M)$ [Eq.\ (\ref{FE1})] without the magnetic field $B$ and the Zeeman 
energy at the metamagnetic field $B_{\rm m}$, respectively. The dotted curve represents the free energy 
$F_{0}(M)-B_{\rm m}M$, which has two degenerate minima at $M=M_{-}$ and ${\bar M}$.    
}
\label{Fig:FreeEnergy}
\end{center}
\end{figure}

In this section, we discuss the Landau theory for the first-order metamagnetic transition 
on the basis of a uniaxial model for magnetization. 
First of all, let us discuss the validity of using the uniaxial model for discussing the metamagnetic transition. 
If there exists an anisotropy in the magnetic response, we have to introduce three order parameters 
corresponding to each direction as long as we follow the Landau-type theory. However, a condition for 
the metamagnetic transition to occur is generally determined independently of its magnetic field direction.  
This is due to a general concept based on the crystal symmetry. Namely, the crystal structure of UTe$_2$ is 
body-centered orthorhombic and centrosymmetric~\cite{Ran} so that the magnetization  
is in the $b$-direction under the magnetic field in the $b$-direction 
because there is no coupling term such as $M_{a}M_{b}$ or $M_{c}M_{b}$ in the Landau-type free energy  
even when the effect of the spin-orbit interaction is taken into account, in which  
magnetic space and real space are not independent owing to the effect of the spin-orbit interaction. 
Therefore, the presence or absence of the metamagnetic transition can be discussed independently of the 
magnetic field direction, justifying the use of the uniaxial model. Therefore, on the basis of experimental facts 
reported in Refs. \citen{AtsushiMiyake, Knafo, Ran2}, we start with the following 
free energy $F_{0}(M)$, with $M$ being the magnetization in the $b$-direction  
per unit formula of UTe$_2$ without the magnetic field ($B=0$):  
\begin{eqnarray}
F_{0}(M)=aM^{2}-bM^{4}+cM^{6}+\cdots
\label{FE1}
\end{eqnarray}
where the coefficients $a$, $b$, and $c$ are assumed to be positive. Near $B=0$, the magnetization is given by 
\begin{eqnarray}
M\approx \frac{B}{2a}\equiv \chi B,
\label{FE2}
\end{eqnarray}
where $\chi$ is the magnetic susceptibility. The stationary condition under a general situation is given by 
\begin{eqnarray}
0=\frac{\partial}{\partial M}[F_{0}(M)-BM]\approx 2aM-4bM^{3}+6cM^{5}-B.
\label{FE3}
\end{eqnarray} 
When the first-order metamagnetic transition occurs, the free energy has at least two degenerate local minima at 
$M=M_{-}$ and ${\bar M}$ as shown schematically in Fig.\ \ref{Fig:FreeEnergy}.  
Therefore, the magnetization ${\bar M}$ at the metamagnetic critical point at $B=B_{\rm m}$ 
is given by the following two conditions: 
The stationary condition of the local minimum of $[F_{0}(M)-B_{\rm m}M]$ for the lower magnetization 
$M_{-}$ at $B=B_{\rm m}$ is 
\begin{eqnarray}
2aM_{-}-4bM_{-}^{3}+6cM_{-}^{5}-B_{\rm m}\approx 0,
\label{FE4}
\end{eqnarray} 
and that for the higher magnetization  ${\bar M}$ at $B=B_{\rm m}$ is 
\begin{eqnarray}
2a{\bar M}-4b{\bar M}^{3}+6c{\bar M}^{5}-B_{\rm m}\approx 0. 
\label{FE5}
\end{eqnarray} 
In addition to these two conditions, the free energy of these two states should be the same: 
\begin{eqnarray}
a{\bar M}^{2}-b{\bar M}^{4}+c{\bar M}^{6}-B_{\rm m}{\bar M}\approx
aM_{-}^{2}-bM_{-}^{4}+cM_{-}^{6}-B_{\rm m}M_{-}.
\label{FE6}
\end{eqnarray}

Note that the anisotropy of a differential magnetic susceptibility around $B=0$ and that of the direction 
in which the metamagnetic transition occurs are generally independent, as noted above.   
This is because the occurrence of the metamagnetic 
transition is determined by the combination of the coefficients $a$, $b$ and $c$ of the extended Landau free energy 
$F_{0}(M)$ [Eq. (1)], whereas the anisotropy in the magnetic susceptibility is determined only by the size of coefficient 
$a$ in each direction of the magnetic field. In this study, we did not touch on the origin of the anisotropy 
in the magnetic susceptibility, but it is left for a future study.

Substituting $B_{\rm m}$ given by Eq.\ (\ref{FE4}) into the r.h.s. of Eq.\ (\ref{FE6}), and 
$B_{\rm m}$ given by Eq.\ (\ref{FE5}) into the l.h.s. of Eq.\ (\ref{FE6}), 
%approximate  expression of  $B_{\rm m}$ [obtained from Eq.\ (\ref{FE4})] into the l.h.s. of Eq.\ (\ref{FE6}), 
we obtain the relation between ${\bar M}$ and $M_{-}$ as 
\begin{eqnarray}
a{\bar M}^{2}-3b{\bar M}^{4}+5c{\bar M}^{6}\approx a{M_{-}}^{2}-3b{M_{-}}^{4}+5c{M_{-}}^{6}. 
\label{FE7}
\end{eqnarray} 
By introducing the parameter $\delta$ defined as $\delta\equiv M_{-}/{\bar M}$, we reduce  Eq.\ (\ref{FE7}) 
to the equation for ${\bar M}$ as 
\begin{eqnarray}
a^{*}{\bar M}^{2}-3b^{*}{\bar M}^{4}+5c^{*}{\bar M}^{6}\approx 0,
\label{FE8}
\end{eqnarray} 
where $a^{*}\equiv a(1-\delta^{2})$, $b^{*}\equiv b(1-\delta^{4})$, and $c^{*}\equiv c(1-\delta^{6})$. 
Solving Eq.\ (\ref{FE8}), we obtain ${\bar M}^{2}$ for the local minimum of the free energy 
$[F_{0}(M)-B_{\rm m}M]$ as 
\begin{eqnarray}
{\bar M}^{2}\approx\frac{3b^{*}+\sqrt{9b^{*2}-20a^{*}c^{*}}}{10c^{*}}. 
\label{FE9}
\end{eqnarray} 

Hereafter, to simplify the analysis, 
we search the case in which there exists a local minimum of $F_{0}(M)$ at $M=M_{0}$, although this 
is not a necessary condition for the occurrence of the metamagnetic transition but slightly narrows the 
parameter space explored. 
Then, there arises a constraint for the coefficient $b$ obtained from the condition $F_{0}(M_{0})>0$, where 
$M_{0}$ is the magnetization corresponding to the stationary condition $\partial F_{0}(M_{0})/\partial M_{0}=0$.  
%for the local minimum of $F_{0}(M_{0})$. 
The stationary condition for $M_{0}>0$ is explicitly given as 
\begin{eqnarray}
a-2bM_{0}^{2}+3cM_{0}^{4}=0,
\label{FE10}
\end{eqnarray} 
so that $M_{0}^{2}$ for the local minimum is given by  
\begin{eqnarray}
M_{0}^{2}=\frac{b+\sqrt{b^{2}-3ac}}{3c}.
\label{FE11}
\end{eqnarray} 
Therefore, with the use of the relation Eq.\ (\ref{FE10}), the condition for $F(M_{0})>0$ is given explicitly as 
\begin{eqnarray}
F_{0}(M_{0})=aM_{0}^{2}-bM_{0}^{4}+cM_{0}^{6}
=\frac{M_{0}^{2}}{3}(2a-bM_{0}^{2})>0.
\label{FE12}
\end{eqnarray} 
By a straightforward calculation, we find that the inequality $2a>bM_{0}^{2}$ gives  the following condition: 
\begin{eqnarray}
3ac<b^{2}<4ac. 
\label{FE13}
\end{eqnarray}

%%%%%%%%%%%%%%%%%%%%%%%%%%%%%%%%%%%%%%%%%%%%%%%%%%%%%%%%%%%%%%%%%%%%%%%%%%%%%%%%%%
\section{Effective Mass and Damping Rate of Quasiparticles at Metamagnetic Transition}
%%%%%%%%%%%%%%%%%%%%%%%%%%%%%%%%%%%%%%%%%%%%%%%%%%%%%%%%%%%%%%%%%%%%%%%%%%%%%%%%%% 
\subsection{Forms of spin fluctuation propagator around two local minima of free energy at metamagnetic transition} 
Under the assumption that the system is uniform in space at around the first-order metamagnetic transition, 
the enhancements of effective mass, i.e., the Sommerfeld coefficient  $\gamma$, and the damping rate of 
the quasiparticles are given by estimating the effect of ferromagnetic spin fluctuations 
in the $b$-direction. Namely, hereafter, we discuss the effect of the longitudinal spin fluctuations
in the $b$-direction, which would be justified by the fact that the enhanced magnetic fluctuations occur 
along the metamagnetic field around $M=M_{-}$ as shown below. 
Note that the transverse spin fluctuations in the $a$- and $c$-directions are essentially unaltered 
because the magnetic field in the $b$-direction does not affect the magnetization perpendicular to the 
$b$-direction in the orthorhombic and centrosymmetric crystal systems such as UTe$_2$ as mentioned in the first 
paragraph of Sect. 2.   
To discuss the effect of  the magnetic fluctuations, we use the conventional 
form of the dynamical spin susceptibility of ferromagnetic fluctuations discussed 
in Refs.\ \citen{Moriya} and \citen{Hatatani} as 
\begin{eqnarray}
\chi_{\rm s}({\boldsymbol q},{i}\omega_{m})
=\frac{qN_{\rm F}^{*}/C}
{\omega_{\rm s}(q)+|\omega_{m}|}, \quad{\hbox{for
$q<q_{\rm c}\sim p_{\rm F}$}}, 
\label{SF1}
\end{eqnarray}
where $N_{\rm F}^{*}$ is the density of states (DOS) of the quasiparticles {\it per spin} at the Fermi level, 
which is renormalized only by the {\it local} correlation effect,  
and $\omega_{\rm s}(q)$ is defined as 
\begin{eqnarray}
\omega_{\rm s}(q)\equiv \frac{q}{C}(\eta+Aq^{2}),
\label{SF2}
\end{eqnarray}
where $\eta$ parameterizes the closeness to the ferromagnetic criticality.~\cite{Comment} 

The static susceptibility around $M=0$ at $B=0$ is given by 
\begin{eqnarray}
\chi_{\rm s}(0,0)=\left[\frac{\partial^{2}F_{0}(M)}{\partial M^{2}}\Big|_{M=0}
\right]^{-1}
=\frac{1}{2a}
\equiv\frac{N_{\rm F}^{*}}{\eta_{0}}. 
\label{SF3}
\end{eqnarray}
This is nothing but the relation Eq.\ (\ref{FE2}). 
By generalizing this relation around $M=0$, we obtain the differential spin susceptibility around $M={\bar M}$ at  
$B=B_{m}$ as 
\begin{eqnarray}
\chi_{\rm s}(0,0)=\left[\frac{\partial^{2}F_{0}(M)}{\partial M^{2}}\Big|_{M={\bar M}}
\right]^{-1}
\equiv\frac{N_{\rm F}^{*}}{{\bar \eta}},
\label{SF3A}
\end{eqnarray}
which is based on the fact that the Zeeman term $(-B_{\rm m}M)$ does not contribute to the curvature of the 
free energy. 
Similarly, that around $M=M_{-}$ at $B=B_{\rm m}$ is given by 
\begin{eqnarray}
\chi_{\rm s}(0,0)=\left[\frac{\partial^{2}F_{0}(M)}{\partial M^{2}}\Big|_{M=M_{-}}
\right]^{-1}
\equiv\frac{N_{\rm F}^{*}}{\eta_{-}}. 
\label{SF3B}
\end{eqnarray}
 
According to the expression for $F_{0}(M)$ [Eq.\ (\ref{FE1})], 
$\partial^{2}F_{0}(M)/\partial M^{2}|_{M={\bar M}}$ is given by 
\begin{eqnarray}
\frac{\partial^{2}F_{0}(M)}{\partial M^{2}}\Big|_{M={\bar M}}
=2a-12b{\bar M}^{2}+30c{\bar M}^{4}.
\label{SF4}
\end{eqnarray}
With the use of Eq.\ (\ref{FE8}), this is reduced to 
\begin{eqnarray}
& &
\frac{\partial^{2}F_{0}(M)}{\partial M^{2}}\Big|_{M={\bar M}}
\approx 2a-12b^{*}{\bar M}^{2}-\frac{6c}{c^{*}}(a^{*}-3b^{*}{\bar M}^{2})
\nonumber
\\
& &
\qquad
=2\left(a+\frac{3a^{*}c}{c^{*}}-4a^{*}\right)+\left(\frac{3c}{c^{*}}-2\right)
\frac{3}{5}\sqrt{9\kappa^{*}-20}\left[\sqrt{9\kappa^{*}-20}+\sqrt{\kappa^{*}}\right]a^{*}, 
\label{SF5}
\end{eqnarray}
where we have used the relation Eq.\ (\ref{FE9}) and $\kappa^{*}\equiv b^{2}/(a^{*}c^{*})$ for obtaining 
the second equality.  
With the use of the definitions of $a^{*}$, $b^{*}$, and $c^{*}$ [see below Eq.\ (\ref{FE8})], 
\begin{eqnarray}
& &
\left(a+\frac{3a^{*}c}{c^{*}}-4a^{*}\right)=\delta^{2}a+\frac{3a\delta^{6}}{1+\delta^{2}+\delta^{4}}\simeq \delta^{2}a,
\label{SF8A}
\\
& &
\left(\frac{3c}{c^{*}}-2\right)=1+\frac{3\delta^{6}}{1-\delta^{6}}\simeq 1, 
\label{SF8B}
\end{eqnarray} 
which is based on the fact that the experimental value for 
UTe$_2$ 
is $\delta\simeq 0.4$.~\cite{AtsushiMiyake}  Therefore, $\partial^{2}F_{0}(M)/\partial M^{2}|_{M={\bar M}}$ is 
approximately given by 
\begin{eqnarray}
\frac{\partial^{2}F_{0}(M)}{\partial M^{2}}\Big|_{M={\bar M}}
\simeq
2\left[\delta^{2}a+\frac{9}{5}\sqrt{9\kappa^{*}-20}\left(\sqrt{9\kappa^{*}-20}+\sqrt{\kappa^{*}}\right)a^{*}\right]. 
\label{SF8C}
\end{eqnarray}
With the use of the inequality [Eq.\ (\ref{FE13})], the value of $\kappa^{*}$ is restricted in the following region:   
\begin{eqnarray}
\frac{3}{(1-\delta^{2})(1-\delta^{6})}<\kappa^{*}<\frac{4}{(1-\delta^{2})(1-\delta^{6})}. 
\label{SF8D}
\end{eqnarray} 
Considering the smallness of $\delta^{4}\simeq 0.026$ and $\delta^{6}\simeq 0.0041$ for UTe$_2$ because 
$\delta\simeq 0.4$,~\cite{AtsushiMiyake} this restriction [Eq.\ (\ref{SF8D})] is technically given by 
\begin{eqnarray}
3.6\lsim\kappa^{*}\lsim 4.8.
\label{SF8E}
\end{eqnarray} 

On the other hand, $\partial^{2}F_{0}(M)/\partial M^{2}|_{M={M_{-}}}$ is given by 
\begin{eqnarray}
& &
\frac{\partial^{2}F_{0}(M)}{\partial M^{2}}\Big|_{M={{M_{-}}}}
=2a-12b{M_{-}}^{2}+30c{M_{-}}^{4}
\nonumber
\\
& &\qquad\qquad\qquad\quad\,\,\,
=2a-12b\delta^{2}{\bar M}^{2}+30c\delta^{4}{\bar M}^{4}.
\label{SF9}
\end{eqnarray}
With the use of the relation ${\bar M}^{4}\approx(-a^{*}+3b^{*}{\bar M}^{2})/5c^{*}$, given by Eq.\ (\ref{FE8}), and 
the definitions of $a^{*}$, $b^{*}$, and $c^{*}$ [see below Eq.\ (\ref{FE8})],  
$\partial^{2}F_{0}(M)/\partial M^{2}|_{M={M_{-}}}$ is reduced to  
\begin{eqnarray}
& &
\frac{\partial^{2}F_{0}(M)}{\partial M^{2}}\Big|_{M={{M_{-}}}}
\approx
2a\left(1-3\delta^{4}\frac{1-\delta^{2}}{1-\delta^{6}}\right)
-6b\left[2\delta^{2}-\frac{3\delta^{4}(1-\delta^{4})}{1-\delta^{6}}\right]{\bar M}^{2}
\label{SF10}
\\
& &
\qquad\qquad\qquad\quad\,\,\,
\simeq
1.87(a-0.78b{\bar M}^{2}),
\label{SF11}
\end{eqnarray}
where we have used $\delta\simeq 0.4$ 
to obtain the approximate equality of Eq.\ (\ref{SF11}).   
Since $\partial^{2}F_{0}(M)/\partial M^{2}|_{M={M_{-}}}$ should be positive, the coefficients $a$ and $b$ and the upper 
magnetic field ${\bar M}$ should satisfy the following condition: 
\begin{eqnarray}
a-0.78b{\bar M}^{2}\gsim0.
\label{SF12}
\end{eqnarray}
Together with the condition  [Eq.\ (\ref{SF8D})], this gives the constraint for the parameter set of 
$a$, $b$, and $c$, and the upper magnetization at $B=B_{\rm m}$.  

To conclude this subsection, it is crucial to note that the relation Eq.\ (\ref{SF11}) generally implies 
that the curvature of the 
free energy at $M=M_{-}$ is always smaller than $2a$ at around $M=0$ under the ambient condition ($B=0$)l, 
which gives considerably larger ferromagnetic fluctuations leading to the enhancements of the Sommerfeld 
coefficient $\gamma$ and the $A_{\rho}$ coefficient of the $T^{2}$ term in the resistivity than those 
under the ambient condition.  
The effect arising from the fluctuations around $M={\bar M}$ at $B=B_{\rm m}$ is smaller than that under the ambient 
conditions because ${\bar \eta}$ is a few times larger than $\eta_{0}$ under the ambient conditions. For example, if we use the average value for $\kappa^{*}=4.2$ over the possible range of the condition [Eq.\ (\ref{SF8E})], 
${\bar \eta}$ is given by ${\bar \eta}\simeq 3.8\eta_{0}$.  

%%%%%%%%%%%%%%%%%%%%%%%%%%%%%%%%%%%%%%%%%%%%%%%%%%%%%%%%%%%%%%%%%%%%%%%%%%%%%%%%%% 
\subsection{Expressions for enhancements of effective mass and damping rate near first-order 
metamagnetic transition} 
On the basis of the the theoretical framework discussed in the previous subsection and 
the formulae given in Appendices A and B, 
we show that the enhancements of the Sommerfeld coefficient  of the quasiparticles and 
the coefficient $A$ of $T^{2}$ term in the resistivity, reported in Refs.\ \citen{AtsushiMiyake} and \citen{Imajo}, 
can be evaluated semiquantitatively by taking a reasonable set of parameters, a part of which is fixed 
using the physical quantities experimentally observed.  

First of all, the magnetization ratio $\delta=M_{-}/{\bar M}$ at the metamagnetic transition is fixed as 
$\delta\simeq 0.4$,~\cite{AtsushiMiyake} 
which has already been used in the previous subsection. The upper magnetization at the 
metamagnetic transition is also determined as ${\bar M}\simeq 1.0$.~\cite{AtsushiMiyake}  
Therefore, the condition [Eq.\ \ref{SF12}] is reduced to 
\begin{eqnarray}
a-0.78b\gsim0.
\label{SF13}
\end{eqnarray}

With the use of Eqs.\ (\ref{SF3A}) and (\ref{SF8C}), ${\bar \eta}$ is given by 
\begin{eqnarray}
{\bar \eta}=N_{\rm F}^{*}
\frac{\partial^{2}F_{0}(M)}{\partial M^{2}}\Big|_{M={\bar M}}
\simeq
\left[2\delta^{2}a+\frac{9}{5}\sqrt{9\kappa^{*}-20}\left(\sqrt{9\kappa^{*}-20}+\sqrt{\kappa^{*}}\right)a^{*}\right]
N_{\rm F}^{*}. 
\label{SF14}
\end{eqnarray}
Similarly, with the use of Eqs.\ (\ref{SF3B}) and (\ref{SF11}), $\eta_{-}$ is given by 
\begin{eqnarray}
\eta_{-}=N_{\rm F}^{*}
\frac{\partial^{2}F_{0}(M)}{\partial M^{2}}\Big|_{M=M_{-}}
\simeq
1.87(a-0.78b{\bar M}^{2})N_{\rm F}^{*}. 
\label{SF15}
\end{eqnarray}
Note that, according to Eq.\ (\ref{SF3}), the parameter $\eta_{0}$ for the ambient condition ($B=0$) is given by 
\begin{eqnarray}
\eta_{0}=2aN_{\rm F}^{*}.
\label{SF3X}
\end{eqnarray}

According to Eqs. (\ref{chiv6A}) and (\ref{B4}), 
the coefficient $A_{\rho}$ of the $T^{2}$ term in the resistivity is given as  
\begin{equation}
A_{\rho}\simeq A_{\rho 0}+\frac{m^{*}}{Ne^{2}}
\frac{VN_{\rm F}^{*}g^{2}C}{4\pi^{2}Nv^{*}_{\rm F}k_{\rm F}A}\left(\frac{1}{\eta}-\frac{1}{\eta+Aq_{\rm c}^{2}}\right),
\label{A_rho}
\end{equation}
where $A_{\rho 0}$ is the coefficient without ferromagnetic spin fluctuations and is given by 
\begin{equation}
A_{\rho 0}\approx\frac{m^{*}}{Ne^{2}}\frac{2}{\epsilon_{\rm F}^{*}}=\frac{m^{*}}{Ne^{2}}\frac{8}{3}N_{\rm F}^{*},
\label{A_rho0}
\end{equation}
where the damping rate $1/2\tau^{*}$ of quasiparticles at the Fermi level is assumed to be given by 
$1/2\tau^{*}\approx sT^{2}/\epsilon_{\rm F}^{*}=(4/3)sN_{\rm F}^{*}T^{2}$, with $\epsilon_{\rm F}^{*}$ 
being the Fermi energy of quasiparticles and $s$ being a constant of ${\cal O}(1)$. 
In deriving the equality in Eq.\ (\ref{A_rho0}), the relation $N_{\rm F}^{*}=(3/4)\epsilon_{\rm F}^{*}$ has been used.  
Note that the second term in Eq.\ (\ref{A_rho}) is not given by 
$\Sigma^{\prime\prime}_{k_{\rm F}}(0;T)$ [Eq.\ (\ref{chiv6})], 
but is given by $\Sigma^{\prime\prime}_{{\rm tr},k_{\rm F}}(0;T)$ [Eq.\ (\ref{chiv6A})].   
Note also that the mass $m^{*}$ in the Fermi energy $\epsilon_{\rm F}^{*}$ appearing in 
Eqs.\ (\ref{A_rho}) and (\ref{A_rho0}), and formulae hereafter, 
is the effective mass 
renormalized both by local correlations among 4f electrons at the U site 
and the effect of spin fluctuations both longitudinal (parallel to $b$-axis) and 
transverse (perpendicular to $b$-axis) under the ambient condition (at $B=0$). 

Similarly, with the use of Eq.\ (\ref{chiv9}), the increase in 
the Sommerfeld coefficient  
due to the ferromagnetic  spin fluctuations, 
$\partial\Sigma_{\rm s}^{\rm R}(p_{\rm F},\epsilon)/\partial\epsilon|_{\epsilon=0}$, leads to 
\begin{eqnarray}
\gamma\simeq \gamma_{0}\left\{1+\frac{VN_{\rm F}^{*}g^{2}}{8\pi^{2}Nv^{*}_{\rm F}A}
\left[\log \frac{Aq_{\rm c}^{2}+\eta}{\eta}
-\frac{1}{2}\log\frac{Aq_{\rm c}^{2}+(Cv^{*}_{\rm F})^{2}}{\eta^{2}+(Cv^{*}_{\rm F})^{2}}\right]\right\},
\label{SF16}
\end{eqnarray}
where $\gamma_{0}\equiv(2\pi^{2}/3)N_{\rm F}^{*}$ is the Sommerfeld coefficient  without the effect of ferromagnetic 
spin fluctuations or renormalized only by {\it local} correlations. 

The coefficients of $VN_{\rm F}^{*}g^{2}C/4\pi^{2}Nv^{*}_{\rm F}k_{\rm F}A$ in Eq.\ (\ref{A_rho}) and 
$Vg^{2}N_{\rm F}^{*}/8\pi^{2}Nv^{*}_{\rm F}A$ in Eq.\ (\ref{SF16}) are estimated as 
\begin{equation}
\frac{VN_{\rm F}^{*}g^{2}C}{4\pi^{2}Nv^{*}_{\rm F}k_{\rm F}A}\approx
\frac{3N_{\rm F}^{*}(Cv^{*}_{\rm F})g^{2}}{16(Ak_{\rm F}^{2})\epsilon_{\rm F}^{*2}},
\label{SF17A}
\end{equation}
and 
\begin{eqnarray}
\frac{VN_{\rm F}^{*}g^{2}}{8\pi^{2}Nv^{*}_{\rm F}A}\approx  \frac{3N_{\rm F}^{*}g^{2}}{16(Ak_{\rm F}^{2})\epsilon_{\rm F}^{*}},
\label{SF17B}
\end{eqnarray}
respectively. In deriving Eqs. (\ref{SF17A}) and  (\ref{SF17B}), we have assumed that the dispersion of 
quasiparticles is given by the free dispersion, i.e., $\epsilon_{\boldsymbol k}=k^{2}/2m^{*}$, so that 
$v^{*}_{\rm F}k_{\rm F}=2\epsilon_{\rm F}^{*}$ and $k_{\rm F}^{3}=3\pi^{2}N/V$. 

In an analysis below, we estimate the coupling constant as $g=4\epsilon_{\rm F}^{*}$ borrowing the theoretical 
result for effective interaction $U^{*}=4T_{\rm K}$ in the single-impurity Anderson model in the Kondo limit.~\cite{Jichu, Yamada}  
For simplicity, we assume $Ak_{\rm F}^{2}\approx Aq_{\rm c}^{2}\approx 1$ 
and $Cv^{*}_{\rm F}\approx 1$. 
These approximations affect to some extent the numerical estimate of the increases in the Sommerfeld coefficient  
$\gamma$ and the $A_{\rho}$ coefficient. However, such an uncertainty will be absorbed in an ambiguity 
of taking many other parameters characterizing the system.  On this approximation scheme, the coefficients in 
Eqs.\ (\ref{SF17A}) and (\ref{SF17B}) are reduced to 
\begin{equation}
\frac{VN_{\rm F}^{*}g^{2}C}{4\pi^{2}Nv^{*}_{\rm F}k_{\rm F}A}\approx 3N_{\rm F}^{*},
\label{SF17A2}
\end{equation}
and 
\begin{eqnarray}
\frac{VN_{\rm F}^{*}g^{2}}{8\pi^{2}Nv^{*}_{\rm F}A}\approx \frac{9}{4},
\label{SF17B2}
\end{eqnarray}
respectively.  
It is crucial to note that the factor $N_{\rm F}^{*}$ arises from  
the density of states per quasiparticle (per spin) contained in the expression of the dynamical spin susceptibility 
[Eqs.\ (\ref{SF1}) and (\ref{chiv1})]. 

One might wonder whether the magnetic field dependence of $N_{\rm F}^{*}$ (or $m^{*}$) 
cannot be neglected 
because, at first sight, the conventional relation $N_{\rm F}^{*}=3/4\epsilon_{\rm F}^{*}$ implies that 
$N_{\rm F}^{*}$ is proportional to the effective mass of the quasiparticles, which 
is affected by the magnetic field of $B_{\rm m}\sim 35\,$T.~\cite{AtsushiMiyake}
However the size of the Zeeman energy (per formula unit) is estimated as 
$\mu_{\rm eff}B_{\rm m}/k_{\rm B}\simeq 9.2\,$K,  
with the effective magnetic moment $\mu_{\rm eff}\simeq 0.4\mu_{\rm B}$ at the metamagnetic field $B=B_{\rm m}$, 
which is  only about 1/5 
of the so-called Kondo temperature $T_{\rm K}$ of $\sim 47\,$K that is estimated from the Sommerfeld 
coefficient under the ambient conditions, i.e., $\gamma_{0}\simeq 1.2\times 10^{2}\,$mJ/K$^{2}$mole\cite{Imajo} 
by comparing with $\gamma\simeq 1.6\times10^{3}\,$mJ/K$^{2}$mole in CeCu$_6$ 
whose $T_{\rm K}\simeq 3.5\,$K.~\cite{Satoh} 

Furthermore, there is a chance that the density of states $N_{\rm F}^{*}$ related to the magnetic susceptibility and 
the Sommerfeld coefficient is technically robust against the magnetic field if the {\it local} spin fluctuations, 
which are the origin of mass enhancement under the ambient conditions, originate mainly from the Van Vleck process 
through the renormalized c-f hybridization in the system with the nonmagnetic (singlet)  
crystalline-electric-field (CEF) ground state in the f$^2$ configuration. 
That is, the Zeeman energy of the quasiparticles is essentially given by 
$\sim -\mu_{\rm B}^{2}(N_{\rm F})_{\rm cond}B^{2}$ because the magnetic susceptibility of the quasiparticles, 
$\chi_{\rm quasi}$, is given by that of conduction electrons, $\chi_{\rm cond}$. 
Indeed, such a property was observed in the NMR Knight shift measurements 
of UPt$_3$,~\cite{Tou} and supported by theoretical discussions in Ref.\ \citen{Ikeda,Yotsuhashi1,Yotsuhashi2}.  
Namely, $N_{\rm F}^{*}$ appearing in Eqs.\ (\ref{A_rho0}) and (\ref{SF17A2}), and $\gamma_{0}$ in Eq (\ref{SF16})  
can be relatively robust against the applied magnetic field. 
Although the CEF ground state of UTe$_2$ has not been observed, we assume that this case is realized as a working 
hypothesis.  

Then, finally, the relations Eqs.\ (\ref{A_rho}) and (\ref{SF16}) are respectively reduced to compact forms as 
\begin{equation}
A_{\rho}\approx \frac{m^{*}}{Ne^{2}}N_{\rm F}^{*}
\left[\frac{8}{3}+3\left(\frac{1}{\eta}-\frac{1}{\eta+Aq_{\rm c}^{2}}\right)\right],
\label{SF17XA}
\end{equation}
and 
\begin{eqnarray}
\gamma\approx \frac{2\pi^{2}}{3}N_{\rm F}^{*}\left\{1+\frac{9}{4}
\left[\log \frac{Aq_{\rm c}^{2}+\eta}{\eta}
-\frac{1}{2}\log\frac{Aq_{\rm c}^{2}+(Cv^{*}_{\rm F})^{2}}{\eta^{2}+(Cv^{*}_{\rm F})^{2}}\right]\right\}. 
\label{SF17XB}
\end{eqnarray}

\subsection{Analysis of experiment of UTe$_2$}
On the parameterization discussed in the previous subsection, we focus on understanding the enhancements of 
$\gamma$ and $A_{\rho}$ at the {\it first-order}  metamagnetic transition reported in Refs.\ \citen{Imajo} 
and \citen{Knafo}. 
The ratios of the Sommerfeld coefficient  and the $A_{\rho}$ coefficient at the metamagnetic transition 
to those at the ambient ($B=0$) state are $\gamma(B=B_{\rm m})/\gamma(B=0)\simeq 2.2$,~\cite{Imajo} and 
$[A_{\rho}(B=B_{\rm m})/A_{\rho}(B=0)]^{1/2}=2.3\pm 0.1$~\cite{Knafo}, respectively. 
We have these two experimental data to explain and two adjustable theoretical parameters, 
$\eta_{0}$ [Eq.\ (\ref{SF3})] and $\eta_{-}$ [Eq.\ (\ref{SF15})], 
that essentially affect the enhancements of $\gamma$ and $A_{\rho}$,  
other than fundamental parameters, such as $Aq_{\rm c}^{2}$ and $Cv^{*}_{\rm F}$ in Eq.\ (\ref{SF16}), 
at the ambient states, and experimentally fixed $\delta$ and ${\bar M}$ as mentioned above. 
On the other hand, ${\bar \eta}$ is restricted to a relatively narrow region.  
Namely, the range of  $\kappa^{*}$ is restricted by Eq.\ (\ref{SF8E}) between 3.6 and 4.8 
so that 
the value of ${\bar \eta}$ is restricted in the range 
\begin{eqnarray}
2.85\,\eta_{0}\lsim {\bar \eta}\lsim 4.78\,\eta_{0},
\label{SF18}
\end{eqnarray}
where we have used Eq.\ (\ref{SF14}) and the relation $2aN_{\rm F}^{*}=\eta_{0}$ [Eq.\ (\ref{SF3X})].  
%Therefore, according to the relations [Eqs.\ (\ref{SF3A}) and (\ref{SF8C})], 
%${\bar \eta}$ is restricted in the range
%\begin{eqnarray}
%2.85\,\eta_{0}\lsim {\bar \eta}\lsim 4.78\,\eta_{0}
%\label{SF18}
%\end{eqnarray}
%where we have used Eq.\ (\ref{SF3X}) for $\eta_{0}$. 
This relation implies that the ferromagnetic fluctuations around $M={\bar M}$ are less important than those 
around $M=0$ at the ambient conditions.  
To find one of the possible sets of parameters that reproduce the observed values of enhancements in 
$\gamma(B=B_{\rm m})/\gamma(B=0)$ and $[A_{\rho}(B=B_{\rm m})/A_{\rho}(B=0)]^{1/2}$ mentioned above, 
let us fix $\kappa^{*}=4.2$,  
which is the average in the possible range [Eq.\ (\ref{SF8E})].  
Then, ${\bar \eta}/\eta_{0}$ is fixed as  ${\bar \eta}/\eta_{0}\simeq 3.85$, 
which is nearly equal to the average over its possible range given by Eq.\ (\ref{SF18}). 

On the other hand, the ferromagnetic fluctuations around $M=M_{-}$ give the most dominant contribution 
to the enhancements of $\gamma$ and $A_{{\rho}0}$, because $\eta_{-}$ [Eq.\ (\ref{SF15})], 
with ${\bar M}\simeq 1.0$, can be considerably smaller than $\eta_{0}=2aN_{\rm F}^{*}$ [Eq.\ (\ref{SF3X})]. 
 
Although the numbers of enhancements of $\gamma$ and $A_{\rho}$ given by Eqs.\ (\ref{SF17XA}) and (\ref{SF17XB}) 
depend on those of $Aq_{\rm c}^{2}$ and $Cv^{*}_{\rm F}$ other than the important parameter $\eta$ characterizing 
the strength of ferromagnetic fluctuations, we are interested in the ratio of those values under 
the metamagnetic field and the ambient condition. 
Therefore, we adopt the set  $Aq_{\rm c}^{2}=1$ and $Cv^{*}_{\rm F}=1$,  
considering that such uncertainties are absorbed in some ambiguities for taking $\eta_{0}$ and $\eta_{-}$.   
Furthermore, we adopt an approximation that a combination $[8/3 -3/(\eta+Aq_{\rm c}^{2})]$ 
in the bracket of Eq.\ (\ref{SF17XA}) can be safely neglected compared with $3/\eta$, which is far larger than 
1 near the ferromagnetic critical point as expected in UTe$_2$.   

Then, it is shown by straightforward arithmetic that the experimental values,  
$\gamma(B=B_{\rm m})/\gamma(B=0)\simeq 2.2$~\cite{Imajo} and 
$[A_{\rho}(B=B_{\rm m})/A_{\rho}(B=0)]^{1/2}=2.3\pm 0.1$,~\cite{Knafo} are reproduced by taking, for example, 
$\eta_{0}=1/10$ and $\eta_{-}=1/50$, which leads to 
\begin{eqnarray}
\frac{\gamma(B=B_{\rm m})}{\gamma(B=0)}=\frac{{\bar \gamma}+\gamma_{-}}{\gamma_{0}}\simeq 2.19,
\label{SF19A}
\end{eqnarray}
and 
\begin{eqnarray}
\left[\frac{A_{\rho}(B=B_{\rm m})}{A_{\rho}(B=0)}\right]=
\left[\frac{{\bar A}_{\rho}+A_{\rho}^{-}}{A_{\rho}^{0}}\right]\simeq 2.29, 
\label{SF19B}
\end{eqnarray}
respectively.  
%many other parameters 
%as mentioned above, there exists essentially almost infinite set of such parameters.Therefore, hereafter, 
%we demonstrate an example of set of parameters which approximately reproduce the observed    
%result $\gamma(B=B_{\rm m})/\gamma(B=0)\simeq 1.9$ and 
%$[A_{\rho}(B=B_{\rm m})/A_{\rho}(B=0)]^{1/2}\simeq 2.4$ reported in Ref.\ \citen{Imajo} and Ref.\ \citen{Knafo}, 
Here, $\gamma_{0}$ and $A_{\rho}^{0}$ are those under the ambient conditions, $\gamma_{-}$ and $A_{\rho}^{-}$ are 
those arising from fluctuations with $\eta_{-}$ around $M=M_{-}$, and ${\bar \gamma}$ and ${\bar A}_{\rho}$ are 
those from fluctuations with ${\bar \eta}$ around $M={\bar M}$. These values [Eqs.\ (\ref{SF19A}) and 
(\ref{SF19B})] reproduce the observed values. 

Of course, there are almost infinite sets of parameters $\eta_{0}$, $\eta_{-}$, and ${\bar \eta}$ [or $\kappa^{*}$ 
through Eqs.\ (\ref{SF3A}) and (\ref{SF5})] to reproduce the observed values of 
$\gamma(B=B_{\rm m})/\gamma(B=0)$ and $[A_{\rho}(B=B_{\rm m})/A_{\rho}(B=0)]^{1/2}$ 
other than those we have adopted above. Furthermore, the ambiguity in the above adopted coupling constant 
$g=4\epsilon_{\rm F}^{*}$ between the quasiparticles and the  
ferromagnetic spin fluctuations may affect the choice of the set of  $\eta_{0}$, $\eta_{-}$ and ${\bar \eta}$. 
Nevertheless, it does not considerably affect the result because the quantities in question are the ratios of values 
between those at $B=B_{\rm m}$ and $B=0$.

%%%%%%%%%%%%%%%%%%%%%%%%%%%%%%%%%%%%%%%%%%%%%%%%%%%%%%%%%%%%%%%%%%%%%%%%%%%%%%%%%% 
\section{Conclusion and Perspective}
Motivated by the experimental observation of the sharp enhancements of the Sommerfeld 
coefficient $\gamma$ and the coefficient $A_{\rho}$ of the $T^{2}$ term in the resistivity at the {\it first-order} 
metamagnetic transition in UTe$_2$, we developed the extended Landau theory of the phase transition 
on the basis of  
the {\it uniaxial} model. As a result, the enhanced ferromagnetic spin fluctuations, caused by 
the flattening of the curvature of the free energy around the shifted local minimum, 
give such enhancements of $\gamma$ and $A_{\rho}$ around 
the metamagnetic field $B_{\rm m}$. 
%the metamagnetic transition magnetic moment 
%at/above the metamagnetic transition. In order to support this idea, the measurement of the differential 
%magnetic susceptibility around  is desired. 
However, the anisotropy in the magnetic response has not been discussed in this paper,
although there exists the pronounced anisotropy in  UTe$_2$. The effect of the anisotropy in the magnetic space 
is left for a future study. 

Another interesting aspect of the present theory is that superconductivity is expected to be induced 
at around $B\lsim B_{\rm m}$ 
owing to the enhancement of ferromagnetic spin fluctuations there, which was the origin of the enhancements of 
$\gamma$ and $A_{\rho}$ around $B=B_{\rm m}$. This mechanism may have some relevance to the appearance 
of the re-entrant superconductivity around the metamagnetic field,\cite{Ran2} along with the discussions 
in the case of URhGe given in Refs.\ \citen{AMiyake1} and \citen{AMiyake2}.

\section*{Acknowledgments}
The author is grateful to Jacques Flouquet and Atsushi Miyake for urging him  to consider the present 
problem and for useful information  and comments to the draft of the present paper.  
The author benefited from conversations with Atsushi Tsuruta, who reminded him of 
the effect of the magnetic field on the spin polarization of the quasiparticles in f$^{2}$-based heavy fermion 
metals with singlet crystalline electric field ground state. 
This work was supported by JSPS-KAKENHI (No. 17K05555).

%%%%%%%%%%%%%%%%%%%%%%%%%%%%%%%%%%%%%%%%%%%%%%%%%%%%%%%%%%%%%%%%%%%%%%%%%%%%%%%%%%

%\newpage
\appendix
\section{Estimation of Self-energy due to Ferromagnetic Fluctuations}
In this appendix, the self-energy of the quasiparticles due to the ferromagnetic fluctuations is discussed. 
For this purpose, we adopt an exponentially decaying phenomenological form for the spin fluctuation propagator 
(dynamical spin susceptibility) $\chi_{\rm s}({\boldsymbol q}, {i}\omega_{m})$ in the Matsubara frequency 
representation as 
\begin{eqnarray}
\chi_{\rm s}({\boldsymbol q},{i}\omega_{m})
=\frac{qN_{\rm F}^{*}/C}
{\omega_{\rm s}(q)+|\omega_{m}|}, \quad{\hbox{for $q<q_{\rm c}\sim p_{\rm F}$}}, 
\label{chiv1}
\end{eqnarray}
where $N_{\rm F}^{*}$ is the DOS of the quasiparticles per quasiparticle, 
and $\omega_{\rm s}(q)$ is defined as 
\begin{eqnarray}
\omega_{\rm s}(q)\equiv \frac{q}{C}(\eta+Aq^{2}),
\label{chiv1A}
\end{eqnarray}
where $\eta$ parameterizes the closeness to the ferromagnetic criticality.~\cite{Moriya,Hatatani} 

\begin{figure}[h]
\begin{center}
\rotatebox{0}{\includegraphics[width=0.4\linewidth]{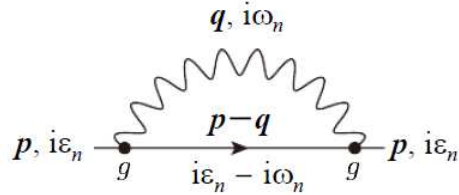}}
\caption{Feynman diagram for the self-energy given by Eqs.\ (\ref{chiv2}) and (\ref{chiv3}). 
The solid line with an arrow represents the Green function ${\bar G}_{\rm f}$ of the f electron renormalized 
only by local correlations, the wavy line represents the spin fluctuation propagator $\chi_{\rm s}$, 
and $g$ is the coupling constant between 
the localized 4f electron and the spin fluctuation mode. }
\label{Fig:1}
\end{center}
\end{figure}

The retarded self-energy $\Sigma_{\rm s}^{\rm R}(p,\epsilon+{\rm
i}\delta)$ gives a measure of the quasiparticle effective mass and
lifetime in its real and imaginary parts, respectively.  It can be
calculated using a simple one-fluctuation mode exchange process
(see Fig.\ \ref{Fig:1}) and is given as 
\begin{eqnarray}
{\rm Re}\Sigma_{\rm s}^{\rm R}(p,\epsilon)&=& -{N_{\rm F}^{*}\over 2\pi CN}\sum_{{\boldsymbol q}}
q\,g_{q}^{2}\int_{-\infty}^{+\infty}{d}x{x\over \omega_{\rm s}(q)^{2}+x^{2}} \nonumber
\\
& &\qquad\qquad\times {\coth{x\over 2T}+\tanh{\xi^{*}_{{\boldsymbol p}-{\boldsymbol
q}}\over 2T}\over -\epsilon+\xi^{*}_{{\boldsymbol p}-{\boldsymbol q}}+x},
\label{chiv2}
\end{eqnarray}
\begin{eqnarray}
{\rm Im}\Sigma_{\rm s}^{\rm R}(p,\epsilon) &=&-{N_{\rm F}^{*}\over
2CN}\sum_{{\boldsymbol q}}q\,g_{q}^{2} {\epsilon-\xi^{*}_{{\boldsymbol p}-{\boldsymbol
q}}\over\omega_{\rm s}(q)^{2}+ (\epsilon-\xi^{*}_{{\boldsymbol p}-{\boldsymbol q}})^{2}}
\nonumber
\\
& &\qquad\quad
\times \biggl(\coth{\epsilon-\xi^{*}_{{\boldsymbol p}-{\boldsymbol q}}\over 2T}
+\tanh{\xi^{*}_{{\boldsymbol p}-{\boldsymbol q}}\over 2T}\biggr), 
\label{chiv3}
\end{eqnarray}
where $N$ is the number of U sites, $g_{q}$ is the coupling between quasiparticles and 
spin fluctuation modes, and $\xi^{*}_{p}$ is the dispersion of the
quasiparticle measured from the chemical potential. 
Hereafter, for simplicity, $g_{q}$ is assumed to be constant
without wavenumber dependence. because it can be essentially approximated as a constant.

In typical limiting cases, (\ref{chiv3}) can be straightforwardly
calculated on the approximation, $\xi^{*}_{{\boldsymbol p}-{\boldsymbol q}}\simeq
-v^{*}_{\rm F}qx$, where $x\equiv\cos\theta$ with $\theta$ being the angle between ${\boldsymbol p}$ and
${\boldsymbol q}$ and $v^{*}_{\rm F}$ being the velocity of the quasiparticles at the Fermi level.  

In the case $T=0$, $\epsilon\not=0$, 
\begin{equation}
{\rm Im}\Sigma_{\rm s}^{\rm R}( p_{\rm F} ,\epsilon)\simeq
-{Vg^{2}N_{\rm F}^{*}C\over32\pi Nv^{*}_{\rm F}\sqrt{A}}
\frac{1}{\eta^{3/2}}\epsilon^{2},
\label{chiv4}
\end{equation}
where $V$ is the system volume, and 
we have used the fact that the factor $\{\coth[(\epsilon-\xi^{*}_{{\boldsymbol p}-{\boldsymbol q}})/2T]
+\tanh(\xi^{*}_{{\boldsymbol p}-{\boldsymbol q}}/2T)\}$ is nonvanishing only in the region 
$0<\xi^{*}_{{\boldsymbol p}-{\boldsymbol q}}<\epsilon$, assuming the case $\epsilon>0$, 
and performed integrations with respect to $q\equiv |{\boldsymbol q}|$ and 
$x\equiv ({\hat {\boldsymbol p}}\cdot{\hat {\boldsymbol q}})$. Here, we have retained the most divergent term 
in $1/\eta$ with $\eta\to 0$ and assumed $\eta\ll Aq_{\rm c}^{2}$.

It is shown by a straightforward calculation that the coefficient of the $\epsilon^{2}$ term for 
the resistivity, ${\rm Im}[\Sigma_{\rm s}^{\rm R}( p_{\rm F},\epsilon)]_{\rm tr}$, 
is less singular owing to the extra factor 
$(q/k_{\rm F})[1-({\hat {\boldsymbol p}}\cdot{\hat {\boldsymbol q}})]=(q/k_{\rm F})(1-x)$, 
which is necessary for taking into account the effect of the momentum change 
${\boldsymbol p}\to{\boldsymbol p}+{\boldsymbol q}$ contributing to  the resistivity. 
Namely, 
\begin{eqnarray}
& &
{\rm Im}[\Sigma_{\rm s}^{\rm R}( p_{\rm F} ,\epsilon)]_{\rm tr}\simeq
-{Vg^{2}N_{\rm F}^{*}C\over16\pi^{2}Nv^{*}_{\rm F}k_{\rm F}A}
\frac{\epsilon^{2}}{\eta}
+{Vg^{2}N_{\rm F}^{*}C\over12\pi^{2}Nv_{\rm F}^{*2}}
\frac{\epsilon^{3}}{(\eta+Aq^{2}_{\rm c})^{2}}
\log\frac{q_{\rm c}\eta}{C\epsilon}
\nonumber
\\
& &
\qquad\qquad\qquad\qquad
-{Vg^{2}N_{\rm F}^{*}C\over32\pi Nv_{\rm F}^{*2}k_{\rm F}\sqrt{A}}\frac{{\epsilon^{3}}}{{\eta}^{3/2}}. 
\label{chiv4A}
\end{eqnarray}

In the case $\epsilon=0$, $0<T\ll\epsilon_{\rm F}^{*}$, 
\begin{eqnarray}
{\rm Im}\Sigma_{\rm s}^{\rm R}(p_{\rm F},0)&\simeq&
-\frac{Vg^{2}N_{\rm F}^{*}}{8\pi^{2} Nv^{*}_{\rm F}C} \int_{0}^{q_{\rm c}}{d}q\,q^{2}
\int_{-vq/T}^{vq/T}{d}y \times\nonumber\\
& & \qquad\quad
{y \over [\omega_{\rm s}(q)/T)^{2}+ y^{2}]}\left(\coth{ y \over 2}-\tanh{ y \over2}\right) 
\label{chiv5},
\end{eqnarray}
where $y=v^{*}_{\rm F}qx/2T$. The integration with respect to $y$ can be approximately 
performed, leading to
\begin{equation}
{\rm Im}\Sigma_{\rm s}^{\rm R}(p_{\rm F},0)\simeq
-{Vg^{2}N_{\rm F}^{*}C \over 8\pi Nv^{*}_{\rm F}\sqrt{A}} 
\frac{1}{\eta^{3/2}}T^{2},
\label{chiv6}
\end{equation}
%\end{document}
where we have made the approximation that the range of integration is technically 
restricted as $-1<y<1$, in which the last factor in (\ref{chiv5})
is approximated as $2/y$.  This is because the factor $(\coth\frac{y}{2}-\tanh\frac{y}{2})$ in Eq.\ (\ref{chiv5}) 
decreases rapidly in proportion to $e^{-y}$ in the region $|y|>1$, and 
$v^{*}_{\rm F}q\gg T$ holds in the dominant region of q-space.  
Similarly to the case $T=0$, $\epsilon\not=0$ above, the imaginary part of the self-energy for the resistivity, 
${\rm Im}[\Sigma_{\rm s}^{\rm R}(p_{\rm F},0)]_{\rm tr}$, is given as 
\begin{equation}
{\rm Im}[\Sigma_{\rm s}^{\rm R}(p_{\rm F},0)]_{\rm tr}\simeq
-{Vg^{2}N_{\rm F}^{*}C \over 4\pi^{2} Nv^{*}_{\rm F}k_{\rm F}A} 
\left(\frac{1}{\eta}-\frac{1}{\eta+Aq_{\rm c}^{2}}\right)T^{2}.
\label{chiv6A}
\end{equation}
Note that $x$ in the extra factor $(q/k_{\rm F})(1-x)$ gives no extra contribution because the integrand 
with respect to $y=v^{*}_{\rm F}qx/2T$ in Eq.\ (\ref{chiv5}) is an even function of $y$. 
%\end{document}

The real part of the self-energy, (\ref{chiv2}), can be calculated easily at $T=0$
and $\epsilon\sim 0$, leading to
\begin{eqnarray}
& &
{\rm Re}\left[\Sigma_{\rm s}^{\rm R}(p_{\rm F},\epsilon)-\Sigma_{\rm s}^{\rm R}(p_{\rm F},0)\right]
\nonumber
\\
& &
\qquad
\simeq
-\epsilon\,\frac{Vg^{2}N_{\rm F}^{*}}{4\pi^{2} Nv^{*}_{\rm F}}\int_{0}^{q_{\rm c}}{d}q\,q
\frac{1}{\eta+Aq^{2}}\left[
1-\frac{(\eta+Aq^{2})^{2}}{(Cv^{*}_{\rm F})^{2}+(\eta+Aq^{2})^{2}}
\right],
\label{chiv7}
\end{eqnarray}
where we have put the external momentum on the Fermi surface, and we also used the following approximate relations 
$\xi^{*}_{{\boldsymbol p}-{\boldsymbol q}}\simeq -v^{*}_{\rm F}qx$ and 
\begin{eqnarray}
& &
\sum_{\boldsymbol q}q\int_{-\infty}^{\infty}{d}w\frac{w}{w^{2}+[\omega_{\rm s}(q)]^{2}}
[{\rm sign}(w)+{\rm sign}(\xi^{*}_{{\boldsymbol p}-{\boldsymbol q}})]
\left(\frac{1}{-\epsilon+\xi^{*}_{{\boldsymbol p}-{\boldsymbol q}}+w}-
\frac{1}{\xi^{*}_{{\boldsymbol p}-{\boldsymbol q}}+w}
\right)
\nonumber
\\
& &
\qquad\quad
\approx
\frac{\epsilon V}{2\pi}\int_{0}^{q_{\rm c}}{d}qq^{3}\frac{1}{v^{*}_{\rm F}q\omega_{\rm s}(q)}
\left\{
1-\frac{[\omega_{\rm s}(q)]^{2}}{(v^{*}_{\rm F}q)^{2}+[\omega_{\rm s}(q)]^{2}}
\right\}+{\cal O}(\epsilon^{2}).
\label{chiv8}
\end{eqnarray}
Performing the ${\boldsymbol q}$-integration, we obtain
\begin{eqnarray}
& &
{\rm Re}\left[\Sigma_{\rm s}^{\rm R}(p_{\rm F},\epsilon)-\Sigma_{\rm s}^{\rm R}(p_{\rm F},0)\right]
\nonumber
\\
& &
\qquad
\approx
-\epsilon\,\frac{Vg^{2}N_{\rm F}^{*}}{8\pi^{2} Nv^{*}_{\rm F}A}
\left[
\log \frac{Aq_{\rm c}^{2}+\eta}{\eta}
-\frac{1}{2}\log\frac{Aq_{\rm c}^{2}+(Cv^{*}_{\rm F})^{2}}{\eta^{2}+(Cv^{*}_{\rm F})^{2}}
\right]+{\cal O}(\epsilon^{2}).
\label{chiv9}
\end{eqnarray}
%\end{document}
Note that the less singular terms in $\eta$ in the logarithm have been retained. 
%\end{document}

\section {Structure of Green Function of Quasiparticles}
In this appendix, we briefly recapitulate the discussion on the relationship between the resistivity and the self-energy 
of quasiparticles, which are strongly renormalized by the self-energy. Let us start with the general expression of 
the retarded Green function $G^{\rm R}({\boldsymbol k},\epsilon)$: 
\begin{eqnarray}
[G^{\rm R}({\boldsymbol k},\epsilon)]^{-1}=\epsilon-\xi^{*}_{\boldsymbol k}
-\Sigma^{\prime}_{\boldsymbol k}(\epsilon)
-{i}\Sigma^{\prime\prime}_{\boldsymbol k}(\epsilon), 
\label{B1}
\end{eqnarray}
where $\Sigma^{\prime}_{\boldsymbol k}(\epsilon)$ and $\Sigma^{\prime\prime}_{\boldsymbol k}(\epsilon)$ are 
the real and imaginary parts of the self-energy, respectively. In the region $\epsilon \sim 0$, 
$[G^{\rm R}({\boldsymbol k},\epsilon)]^{-1}$ is approximated as 
\begin {eqnarray}
& &
[G^{\rm R}({\boldsymbol k},\epsilon)]^{-1}\simeq 
\left[1-\frac{\partial \Sigma^{\prime}_{\boldsymbol k}(\epsilon)}{\partial\epsilon}\right]_{\epsilon=0}\times
\nonumber \\
& &\qquad\qquad\quad
\left\{\epsilon
-\left[1-\frac{\partial \Sigma^{\prime}_{\boldsymbol k}(\epsilon)}{\partial\epsilon}\right]^{-1}_{\epsilon=0}
\xi^{*}_{\boldsymbol k}
%-\Sigma^{\prime}_{\boldsymbol k}(\epsilon)
-{i}\left[1-\frac{\partial \Sigma^{\prime}_{\boldsymbol k}\partial(\epsilon)}{\partial\epsilon}\right]^{-1}_{\epsilon=0}
\Sigma^{\prime\prime}_{\boldsymbol k}(\epsilon)
\right\}.
\label{B2}
\end{eqnarray}
Therefore, the effective mass ${\tilde m}^{*}$ near the Fermi level and the damping rate 
$1/{\tilde \tau}^{*}_{\rm tr}(\epsilon;T)$ with energy $\epsilon$ are renormalized as 
\begin {eqnarray}
{\tilde m}^{*}\simeq\left[1-\frac{\partial \Sigma^{\prime}_{\boldsymbol k}(\epsilon)}{\partial\epsilon}\right]_{\epsilon=0}
m^{*},
\label{B3}
\end{eqnarray}
and 
\begin {eqnarray}
\frac{1}{\tau^{*}_{\rm tr}(\epsilon;T)}\simeq -
\left[1-\frac{\partial \Sigma^{\prime}_{\boldsymbol k}(\epsilon)}{\partial\epsilon}\right]^{-1}_{\epsilon=0}
\Sigma^{\prime\prime}_{{\rm tr},{\boldsymbol k}}(\epsilon;T),
\label{B4}
\end{eqnarray}
respectively. Here, $\Sigma^{\prime\prime}_{{\rm tr},{\boldsymbol k}}(\epsilon;T)$ is the imaginary part of the 
self-energy in which the effect of the momentum change is taken into account as mentioned in the discussion 
leading to Eq.\ (\ref{chiv4A}).  
Then, it is immediately found that the approximate Drude formula for the resistivity 
$\rho\simeq ({\tilde m}^{*}/Ne^{2})[1/{\tilde \tau}_{\rm tr}^{*}(0;T)]$ is not affected by the renormalization factor 
$ [1-\partial \Sigma^{\prime}_{\boldsymbol k}(\epsilon)/\partial\epsilon]^{-1}_{\epsilon=0}$.
~\cite{MMV,Varma}
Here, we have made the approximation that $\langle1/{\tilde \tau}^{*}_{\rm tr}(\epsilon;T)\rangle$, 
where $\langle\cdots\rangle$ is the average over $\epsilon$ with the weight of 
the $\epsilon$-derivative of the Fermi distribution function [$-\partial f(\epsilon)/\partial\epsilon$], is replaced by 
$1/{\tilde \tau}^{*}_{\rm tr}(0;T)$, considering that $\epsilon$ and $T$ dependences appear through a combination 
of $[\epsilon^{2}+(\pi T)^{2}]$ as in the Fermi liquid state.~\cite{AGD}

Namely, the resistivity $\rho$ is given approximately as 
\begin{eqnarray}
\rho\simeq\frac{{\tilde m}^{*}}{Ne^{2}}\frac{1}{{\tilde \tau}^{*}_{k_{\rm F},\rm tr}(0;T)}
\approx\frac{m^{*}}{Ne^{2}}\left[-\Sigma^{\prime\prime}_{{\rm tr},k_{\rm F}}(0;T)\right]. 
\label{B5}
\end{eqnarray}

%%%%%%%%%%%%%%%%%%%%%%%%%%%%%%%%%%%%%%%%%%%%%%%%%%%%%%%%%%%%%%%%%%%%%%%%%%%%%%%%%%


\begin{thebibliography}{99}
\bibitem{Comment}
Note that this parameterization is also valid even in the case where the ground state is a 
non-ferromagnetic state.  Indeed, it has been successfully explained 
by the self-consistent renormalization (SCR) theory that the Curie-Weiss law for the static susceptibility 
is obtained down to the low-temperature region ($T\ll T_{\rm F}$) even in metallic systems, 
i.e., without the localized magnetic moment, which is consistent with the results of 
a series of experiments on some metallic and magnetic systems.~\cite{Moriya, Hatatani}  
However, this SCR effect gives higher-order correction  in the low temperature dependence 
in the non-ferromagnetic state such as $C/T$ and $A$ coefficient.
\end{thebibliography}
\end{document}